# Child labour and schooling decision of the marginal farmer households: An empirical evidence from the East Medinipur district of West Bengal, India

Sangita Das

Faculty Member, Syamaprasad College, Department of Economics

Kolkata-700025, West Bengal, India

## Abstract

Based on the field investigation of West Bengal, this paper investigates whether the school-aged children of the marginal farmer households are full-time paid labourers or unpaid domestic labourers along with schooling or regular students. Probit Regression analysis is applied here to assess the influencing factors for reducing the size of the child labour force in practice. The result shows that the higher is the earning of the adult members of the households, the lower is the incidence of child labour. Moreover, the credit accessibility of the mother from the Self-help group and more person-days of the father in work in a reference year are also responsible for reducing the possibility of a child turning into labour. The study further suggests that the younger age of the father, father's education, and low operational landholdings are positive and significant determinants to decide on a child's education by restricting their excessive domestic work burden.

## Introduction

The prevalence of child labour in India is still an acute socio-economic problem. The children who are working under the legal minimum working age[1] of 18 years are considered as child labour. They are forced to discontinue school prematurely, neglected physically, socially, or morally, and unable to explore their potentiality, personality, and talents due to excessive work burden [ILO Minimum Age Convention (1973)]. Although all activities performed by children (below the age of 18) are not treated as child labour, most of the home-based children's work[2] deprives them of their childhood. Child labour in domestic work such as in agriculture, or any non-farm activities or domestic chores even after school hours, are detrimental to their education [Basu (2010) and George & Panda (2015)]. As per the International Labour Organization [ILO (1998A)], if a 5-11-year-old child performs in any economic activity at least one hour a week and the 12-14-year-old child involves at least either 14 hours of non-hazardous work per week or 1 hour of hazardous work per week are regarded as child labour. The share of working children of the total child population[3] of India is 3.9 percent which figures more than 10 million children [Census of India (2011]. A large number of children belonging to marginal farmers or landless labour households are working as main or marginal agricultural workers of their own land and (or) leased-in land [George & Panda (2015)]. In India, around 59 percent (5.99 million) of child labour as the proportion of the total child labour (10.13 million) have participated in agriculture either as cultivators (2.63 million) or as agricultural labourers (3.33 million) [Census of India (2011)].

Many researchers have explained several reasons for the existence of child labour. Poverty, lack of educational attainment, and social and cultural stigma are responsible for the high persistence of child workers in India [Basu (1999)]. Agricultural income shock among the poor is another crucial factor for increasing household working hours of children that adversely impacts their school attendance [Bandara and et al. (2014)]. Based on the NFHS individual-level data, Dev (2004) examined the role of socio-economic factors on the chances of labour force participation of children. A logistic regression analysis shows that the possession of cultivable land and livestock in the household enhances the possibility of a child's participation in work. Furthermore, mothers' education, household size, employment diversification, and family income, and female labour force participation rates are the influential determinants of children's schooling-work decision [Burki and et al. (1998) & Sasmal and Guillen (2015)]. Baland & Robinson (2000) argued that child labour is in equilibrium, even if it may not be socially efficient. They explained that child labour would be inefficient if parents compel their children to engage in any economic activities for alternative family income or the imperfect capital market exists in the economy. Basu (2000) developed a theoretical model and examined the impact of minimum wage legislation on child labour. The higher adult wage rate by adopting minimum wage legislation is one of the critical factors for reducing adult labour demand and increasing child labour supply in an economy. Goswami & Jain (2006) discussed the various poverty and non-poverty-related factors as the responsible determinants of child labour in developing countries and their impact on children's welfare. The poverty-related factors such as low per capita income of the household, less accessibility of credit and crop failure, and non-poverty factors such as parental attitude towards child's work, children belonging to scheduled caste, low level of parental education, and male-dominated households affect the household's decision regarding the labour force participation of a child. Based on the primary survey data of Uttaranchal and Himachal Pradesh, Basu and et al. (2010) tried to assess the impact of wealth on child labour. They confirmed theoretically and empirically that the relationship between land ownership and child labour is an inverted U-shaped, i.e., child's work increases at a decreasing rate with the increase in household's possession of landholdings. The labour market imperfections might be the key reason for not obtaining any monotonically declining relationship between wealth and child labour. The paper of Guarcello & at el. (2010) focused on human capital investment for the eradication of the incidence of child labour. Better accessibility of institutional credit helps to reduce the risk and vulnerability of the households that encourage parents to invest in the human capital for the development of their children.

---

[1] Children under the age of 14 years are not to be employed in any hazardous work, factory, or mine [Article 24 of the Constitution of India].

[2] Some works are acceptable for the children such as working with parents or farmers that neither prevents to continue school nor heavy work nor barrier to participate them any social and family activities [ILO & UN convention on the Rights of the Child]. ILO Convention (189) further stated that child domestic work covers all types of permissible (allowed by law) and non-permissible (not allowed by law) activities. ILO (2012) estimated that more than 17 million children (age group 5 and 17 years) are either paid or unpaid domestic workers found in Asia and Sub-Saharan Africa.

[3] As per the Census of India (2011), India's total child population (in the age group of 5-14 years) is 259.6 million.

In India, several schemes have already been introduced to prevent the persistence of child labour and to rehabilitate children engaging in the labour force. The policies such as the National Child Labour Policy (1988), the Child Labour (Prohibition and Regulation) Act (1986), the International Programme on the Elimination of Child Labour (IPEC) in collaboration with ILO (1992), and The Right of Children to Free and Compulsory Education Act (2009) have implemented to provide the right to education to all children, especially, to the unprivileged section of children and protect them from the engagement of any hazardous work and exploitation. Although the number of working children of age group 5-14 years has decreased[4] from 8.57 lakh to 2.34 lakh between 2001 and 2011 (Census data), the persistence of child labour remains a serious social issue in West Bengal. The state has witnessed around 6 percent of the country's total working children [Census of India (2011)]. These child labourers are working for long hours in an exploitative manner, being exposed to toxic chemicals, and suffering high rates of health hazards that violate the Fundamental Human Rights.

In the background, this paper puts an effort to investigate the possible indicators responsible for reducing the incidence of school-going domestic child labour or (and) school dropouts engaged in child labour among the marginal farmer households in the district East Medinipur in West Bengal.

## Data

### *Survey design and field investigation*

Bhagawanpur block-1, one of the economically backward blocks of the East Medinipur district of West Bengal, has been selected purposively for the study. The district contributes 4.1 percent of child labour of the total child labour of West Bengal [Census of India (2001)].

Two large Gram Panchayats (GPs) of the block: Mahammadpur-I and Mahammadpur-II have been considered here. Out of 22 villages of these two GPs, nine villages were chosen randomly. The survey conducted among the marginal farmer households[5] who possessed up to one bigha (or 53.33 decimal) of cultivable land as they are one of the poorer sections of the rural population. Target households are first identified from gram panchayat's recorded list and then selected randomly.

To investigate the research objective, a structured questionnaire has been designed based on the extensive pilot survey of two villages of this block among the marginal farmer households who have at least one child in the age group 5-14 years. The total sample for this investigation is 372 households. As the primary survey was conducted between May and June 2019, the reference period for this field investigation would be from June 2018 to July 2019.

### *Data Analysis*

**Table-1** Statistics of the Children (in the age group 5-14 Years) engaged in economic activities of the marginal farmer households (in percentage)

| The households who have at least one child in the 5-14 age groups | No. of households | Households (%) |
|---|---|---|

[4] In India, child labour has decreased by 1.1 percent (or 2.6 million) during this period.

[5] These households are observed to cultivate own land and (or) leased-in land and engage in several non-farm diversified activities for improving their living standard.

| School-going children and do not economically active | 192 | 51.61 |
|---|---|---|
| School-going children engaged in economic activity within the family and (or) performed household chores as unpaid domestic workers. | 166 | 44.62 |
| Dropout children before class IX and work at different places (other than their homes) as paid child labourers | 14 | 3.76 |
| Total | 372 | 100 |

Source: Author's calculation from field investigation

Table-1 shows that the status of the children of 5-14 years in the marginal farmer households. The school-going children without involvement of any economic activities are 51.61 percent. In the second case, around 45 percent of children are working[6] along with schooling as unpaid family labourers of these households. Parliamentary Standing Committee on Labour (2013-14) stated that performing domestic chores by school-going children may adversely impact their study as leisure and recreation are required to develop their mental and physical health [George and Panda (2015)]. Some domestic work such as cooking, taking care of younger siblings and, long and exhaustive working hours results in harmful impacts on children's physical and moral development [Basu (2010)]. On the contrary, 3.76 percentage of school drop-out children before class IX are working as paid labourers outside their homes in the study area. They are mainly migrated child labourers and worked in the gems and jewellery industry, hotel/restaurants, or bag/shoe manufacturing factories.

### Investigating the possible factors for reducing the incidence of school dropout paid-child labour outside the home and school-going unpaid child labour within the family of the marginal farmer households

The paper tries to investigate the responsible factors which can reduce the domestic workload on the school-going children, and the incidence of school dropouts engaged as child labour among the marginal farmer households, and the theoretical justifications of these regressors are given below-

1. Father's age of the $i^{th}$ household ($age_i$): The relationship between the aged father and the possibility of working children in the family is assumed to be positive. Sometimes an older father might be unable to earn adequate income during a year. In this case, either school-going children have to involve in economic activities at home for earnings or drop out children join in work as paid labourers to maintain the minimum living standard of the household.

[6] School-going children of the households have participated as family labourers in agriculture and allied activities (such as betel leaf, fisheries, livestock rearing, and vegetable cultivation). They also help other adult family members in non-farm activities (namely hair processing, bidi binding, part-time duty in the local shop, etc.) and perform domestic chores (such as cooking, cleaning, washing, collecting fuelwoods, and taking care of siblings, etc.).

2. Housing infrastructure i[th] households ($houseinfra_i$): In the study region, although 58 percent of households possess two rooms with separate kitchens in their houses, most of these houses (50.6 percent) are semi-pucca houses. Hence, it is required to examine whether housing infrastructure impacts children's labour force participation rate inside and outside their homes.
3. Caste of the households ($caste_i$): A high incidence of child labour is observed in the socially backward class (such as SC, ST, and OBC) households as they are suffering from a high level of poverty than the upper caste households [Goswami & Jain (2006)]. Therefore, caste may play a key role in increasing the working children of these households.
4. Father's education of the i[th] households ($fatheredu_i$): As India's school education up to the class VIII is almost zero, a literate father is assumed to provide their children quality education that helps to develop children's potential for getting a higher return in future.
5. Mother's education of the i[th] households ($motheredu_i$): Literate mother is expected to send their children to school, and encourage, especially, girl children to attend the school regularly [Burki & et al. (1998)].
6. Possession of operating landholdings (in decimal[7]) of the i[th] households ($opeland_i$): Children belonging to the households possessed by operational landholdings (owned and leased-in land) are assumed to involve as the family labourers along with schooling in agriculture and allied activities, especially, the peak harvest season, as labour hiring cost that time is very high [Woldehanna & et al. (2008)].
7. The total number of unemployed days of the father ($unemplydys_i$): It is measured in the number of person-days in a year. The association between high unemployed person-days of the father in the reference period and working children can be positive. Instead of providing school education, the unemployed father probably sends their children to work as the hired labourers for supplementary family income.
8. The number of earning members of the i[th] households ($earnmem_i$): Except for household heads, if other adult members are involved in diversified economic activities that might enhance the possibility of children's schooling instead of working as monthly income of these households is expected to be sufficient to meet their subsistence standard of living.

[7] 1 Bigha land = 53.33 Decimal

9. Mother's accessibility of Self-help group credit of the $i^{th}$ households ($shgcredit_i$): Self-Help Group loans empower rural women to become economically self-reliant by creating employment opportunities that play a key role in the decision-making of their children's well-being.

## Econometric Investigation

To examine the influencing determinants for child's 'labour' and 'labour-schooling' decision of the marginal farmer households, two dependent variables such as school dropouts engaged as paid child labour outside the home (childlabour) and school-going unpaid child labour in domestic work (cldw) are binary in nature, considered in the two separate Probit regression models. The Probit regression equation for model-1 can be expressed as-

$$childlabour_i = \alpha_0 + \alpha_1(age_i) + \alpha_2(houseinfra_i) + \alpha_3(caste_i) + \alpha_4(fatheredu_i) + \alpha_5(motheredu_i) + \alpha_6(opeland_i) + \alpha_7(unemplydys_i) + \alpha_8(earnmem_i) + \alpha_9(shgcredit_i) + u_{1i} \quad \ldots \ldots \ldots (i)$$

The Probit regression equation for model-2 can be expressed as-

$$cldw_i = \beta_0 + \beta_1(age_i) + \beta_2(houseinfra_i) + \beta_3(caste_i) + \beta_4(fatheredu_i) + \beta_5(motheredu_i) + \beta_6(opeland_i) + \beta_7(unemplydys_i) + \beta_8(earnmem_i) + \beta_9(shgcredit_i) + u_{2i} \quad \ldots \ldots \ldots (ii)$$

Now, the descriptive statistics of all variables including two dependent variables used in equation (i) and (ii) are described as in the Table-2.

**Table 2** Descriptive Statistics of the regressors including two outcome variables

| Variables | Variable descriptions | Mean | Std. Dev | Max | Min |
|---|---|---|---|---|---|
| childlabour | =1 if dropouts engaged as child labourers, otherwise (children who are regular student and do not work) = 0 | 0.1542 | 0.0242 | 1 | 0 |
| cldw | =1 if children are in domestic work with schooling, otherwise (children who are regular student and do not work) = 0 | 0.4931 | 0.3734 | 1 | 0 |
| age | Father's age (in years) | 48.2903 | 8.763 | 70 | 26 |
| houseinfra | = 1 if housing and flooring infrastructure is kutcha, otherwise (kutcha-pucca/pucca) = 0 | 0.5806 | 0.4941 | 1 | 0 |
| caste | =1 if the household belongs to SC/ST/OBC; otherwise (General Caste) = 0 | 0.7043 | 0.4569 | 1 | 0 |
| fatheredu | Number of schooling years of the father | 8.2311 | 3.2901 | 17 | 0 |
| motheredu | Number of schooling years of the mother | 6.6612 | 3.7943 | 13 | 0 |
| opeland | Size of operational landholdings (own and/or leased-in land in terms of decimal) for cultivation | 25.23 | 19.2634 | 55 | 0 |
| unemplydys | Number of father's unemployed person-days during the reference year | 25.483 | 20.42 | 180 | 0 |
| earninmem | Total number of earning members including household head of the family | 2.5026 | 0.7469 | 5 | 1 |
| shgcredit | =1 if mother's accessibility of credit from Self-Help Group, otherwise=0 | 0.7526 | 0.43203 | 1 | 0 |

Source: Author's calculation

The results of the two Probit regression models are represented as in the following Table-3.

**Table-3:** Probit Regression Analysis[8]: Two dependent variables: 'childlabour' and 'cldw'

| | Equation (i)<br>Model-1 | Equation (ii)<br>Model-2 |
|---|---|---|
| Observations | 206 | 358 |

[8] There is no multi-collinearity problem in the Probit regression model, as the values of Variance Inflation Factor (VIF) of all explanatory variables are less than 4.

| Variables | Value of coefficients | Average Marginal effect $\left(\frac{\partial p_i}{\partial x_i}\right)$ | Value of coefficients | Average Marginal effect $\left(\frac{\partial p_i}{\partial x_i}\right)$ |
|---|---|---|---|---|
| age | 0.004002 (0.0273) | 0.00016 | 0.00915* (0.0048) | 0.00247* |
| caste | 0.9317** (0.3554) | 0.03767** | 0.0614 (0.16351) | 0.02333 |
| houseinfra | 0.7379 (0.6169) | 0.02983 | 0.1462 (0.1381) | 0.0555 |
| motheredu | 0.0444 (0.0756) | 0.00179 | -0.0051 (0.0215) | -0.0019419 |
| fatheredu | -0.02214 (0.0847) | -0.00089 | -0.0533** (0.02502) | -0.02024** |
| earningmem | -0.6013* (0.3691) | -0.0243* | -0.1523* (0.0771) | -0.0578* |
| umemplydys | 0.03906* (0.00279) | 0.0077* | 0.00242 (0.00303) | 0.00091 |
| land | 0.000325 (0.0037) | 0.00012 | 0.0572* (0.0197) | 0.0015* |
| shgcredit | -1.0782* (0.0263) | -0.0435* | -0.0809 (0.1632) | -0.0307 |
| Constant | 3.497* | - | 1.193* | - |
| Pseudo $R^2$ | 0.3237 | - | 0.1031 | - |
| Chi-square (9) | 15.23 | - | 11.21 | - |

**Note**: *** indicates 1% level of significance, ** indicates 5% level of significance and * indicates 10% level of significance. Standard errors of the coefficients are written in parenthesis.

## Discussion

Table-3 shows that the households belonging to socially backward classes (such as SC/ST/OBC) are responsible for increasing the incidence of child labour among the marginal farmer households. More earning members in the family earn enough income for their living standard and encourage the children to be regular at school. But, loss of income due to high unemployed person-days of the father sometimes obliges the children to engage in economic activity as the hired labourer. Mother's employment opportunity through the accessibility of self-help group loans has a positive and significant impact on children's well-being by reducing

the incidence of child labour. In the second model of Probit Regression Analysis, it is observed that aged father sometimes compels children to work as family labourers along with schooling. Educated fathers always want to pursue children's education even beyond the secondary level to make them productive and skilled labour in the future. The households with the possession of operating landholdings and working children as the family labourers outside school hours are significant and positively associated with each other as most of these households depend on family labour, especially during the peak season of agriculture.

## Conclusion and policy implications

The present study tries to identify the major determinants which can reduce the incidence of child labour inside and outside of the family among the marginal households of the East Medinipur district of West Bengal. The Probit Regression analysis shows that the family with a higher number of earning members prefer to continue the child's education. The family belongs to the general caste and a smaller number of unemployed days of the father have a negative and significant impact on the probability of being full-time child workers. The paper further explains that the mothers' credit accessibility from the Self-help group facilitates them to earn an additional amount of money that supports to continue their children's education and prevent them from being forced to work as child laborers. The second model of Probit Regression Analysis suggests that the fathers' education plays a vital role in continuing their children's education. On the contrary, the possession of landholdings of the households increases the risk of school-going children engaged in farm / non-farm activities as the family labourers.

The study has recommended some policies that might enhance the welfare of children by reducing the incidence of child labour in these households. Parental awareness about the child's education, mother's empowerment through employment opportunity by the accessibility of self-help group loans and a decent standard of living of the households by involving adult family members in several economic activities can reduce the incidence of child labour. In addition, technical and vocational education and training programmes are required to conduct in schools for the students after class VIII that can encourage parents to invest in children's education for their future job prospects. Some awareness programmes must also be launched by the Govt./ NGO to educate parents about how children are exploited, abused, and deprived due to the incidence of child labour and make them understand the usefulness of children's education and its future return.